\documentclass{article}

\usepackage{arxiv}

\usepackage[utf8]{inputenc} 
\usepackage[T1]{fontenc}    
\usepackage{hyperref}       
\usepackage{url}            
\usepackage{booktabs}       
\usepackage{amsfonts}       
\usepackage{nicefrac}       
\usepackage{microtype}      
\usepackage{lipsum}
\usepackage{graphicx}

\title{Pythia: AI-assisted Code Completion System}

\author{
  Alexey Svyatkovskiy \\
  Microsoft\\
  One Microsoft Way\\
  Redmond, WA 98052 \\
  \texttt{alsvyatk@microsoft.com} \\
   \And
  Ying Zhao \\
  Microsoft\\
  One Microsoft Way\\
  Redmond, WA 98052 \\
   \And
  Shengyu Fu \\
  Microsoft\\
  One Microsoft Way\\
  Redmond, WA 98052 \\
    \And
  Neel Sundaresan \\
  Microsoft\\
  One Microsoft Way\\
  Redmond, WA 98052 \\ 
}

\begin{document}
\maketitle

\begin{abstract}
In this paper, we propose a novel end-to-end approach for AI-assisted code completion called Pythia. It generates ranked lists of method and API recommendations which can be used by software developers at edit time. The system is currently deployed as part of Intellicode extension in Visual Studio Code IDE. Pythia exploits state-of-the-art large-scale deep learning models trained on code contexts extracted from abstract syntax trees. It is designed to work at a high throughput predicting the best matching code completions on the order of 100 $ms$. 

We describe the architecture of the system, perform comparisons to frequency-based approach and invocation-based Markov Chain language model, and discuss challenges serving Pythia models on lightweight client devices. 

The offline evaluation results obtained on 2700 Python open source software GitHub repositories show a top-5 accuracy of 92\%, surpassing the baseline models by 20\% averaged over classes, for both intra and cross-project settings. 
\end{abstract}


\section{Introduction}

In software development through Integrated Development
Environments (IDEs)~\cite{murphy2006}, code completion is one of the most
widely used features. Intelligent code completion~\cite{proksch2015intelligent} assists
developers by reducing typographic and other common errors, effectively improving developer productivity. These features may include pop-ups when typing, calling methods and APIs on typed
classes and objects, querying parameters of functions, variable
or function name disambiguation, and program structure
completion that use code context to reliably predict following code.

Traditional code completion tools in IDEs typically list out all possible
attributes or methods that can be invoked when
a user types a "." or an "=" trigger character. However, not ranking the suggested results
requires users to scroll through a long alphabetically ordered
list, which is often even slower than typing the full name
of a method directly. Studies show that developers tend
to rely heavily on prefix filtering to reduce the number of choices \cite{ppig2015}. 

In this paper, we introduce Pythia -- a neural code completion system, trained on code snippets extracted from a large-scale open source code dataset.
The fundamental task of the system is to find
the most likely method given a code snippet. In other words, given original code snippet $C$, the vocabulary $V$, and the set of all possible methods $M \subset V$, we would like to determine:
\begin{equation}
m^{*} = argmax(P(m|C)), \forall m \in M. 
\end{equation}
In order to find that method, we construct a model capable of scoring responses and then determining the most likely method. 

A large number of intelligent code completion systems for both statically and dynamically typed languages have been proposed in the literature \cite{d2016collective,raychev2014code,proksch2015intelligent,bruch2009learning,ALNUSAIR2010,GVERO2013}. Best Matching Neighbor (BMN) and statistical language models such as n-grams, as well as recurrent neural network (RNN) based approaches leveraging sequential nature of the source code have been particularly effective at creating such systems. The majority of the past approaches did not leverage the long-range sequential nature of the source code or tried to transfer natural language methods without taking advantage of unique opportunities offered by code's abstract syntax trees (AST). 

The nature of the problem of code completion makes long short-term memory (LSTM) networks~\cite{lstm} a promising candidate. Pythia consumes partial ASTs corresponding to code snippets containing member access expressions and module function invocations as an input for model training, aiming to capture semantics carried by distant nodes.

The main contributions of the paper are: (i) we introduce and implement several baselines of code completions systems, including frequency-based and Markov Chain models, (cf. section~\ref{sec:baselines}), (ii) we propose and deploy a novel end-to-end code completion system based on LSTM trained on source code snippets (cf. sections~\ref{sec:code2vec} and~\ref{sec:neural}), (iii) we evaluate our model on a dataset of 15.8 million method calls extracted from real-world source code, showing that our best model achieves 92\% accuracy, beating simpler baselines (cf. section~\ref{sec:evaluation}), (iv) we discuss and document practical challenges of training deep neural networks and hyperparameter tuning on high-performance computing clusters, and model deployment on lightweight client devices at high throughput (cf. section~\ref{sec:deploy}).

\section{Baseline code completion systems}
\label{sec:baselines}

\begin{figure*}
    \includegraphics[width=.49\textwidth]{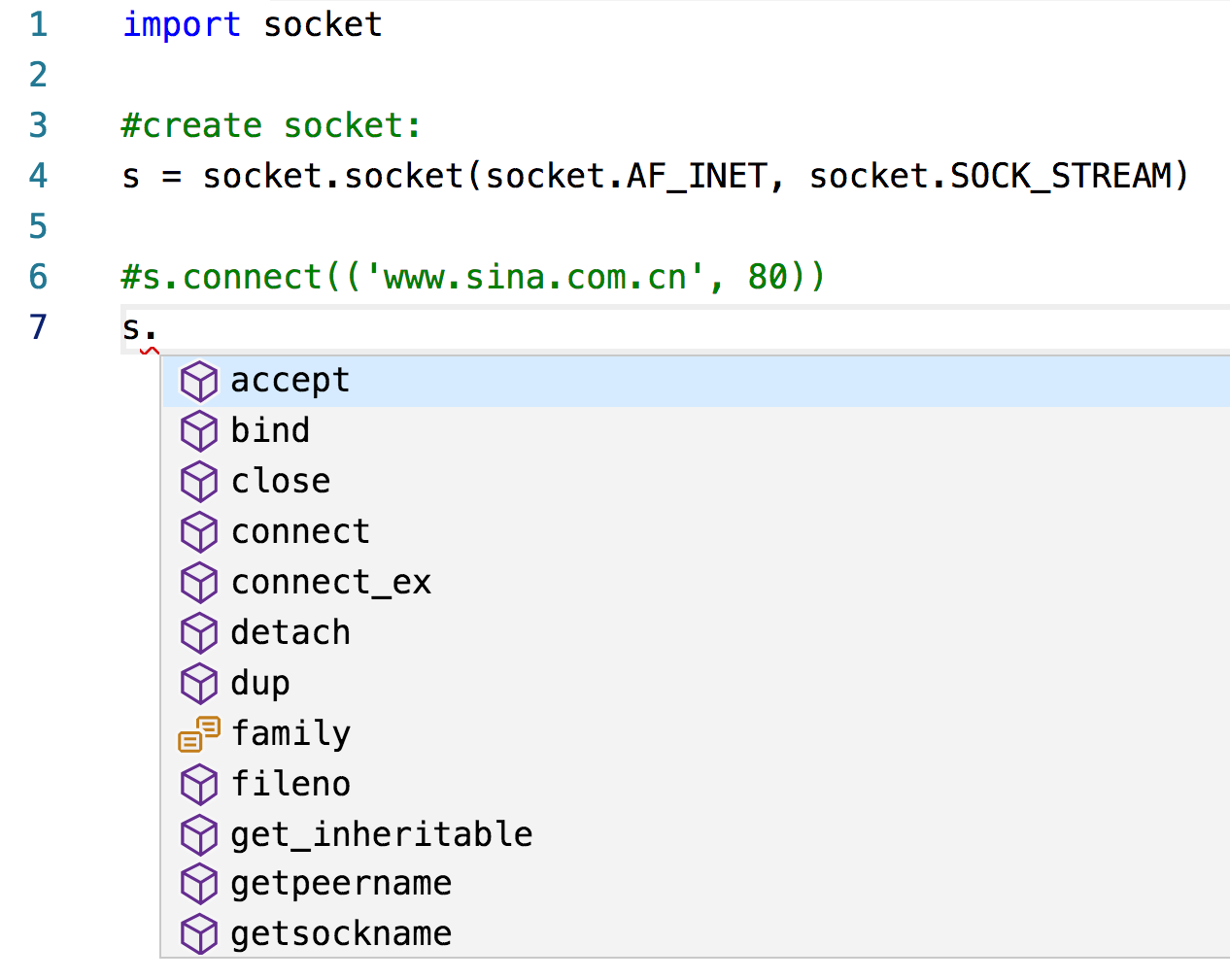}
    \includegraphics[width=.49\textwidth]{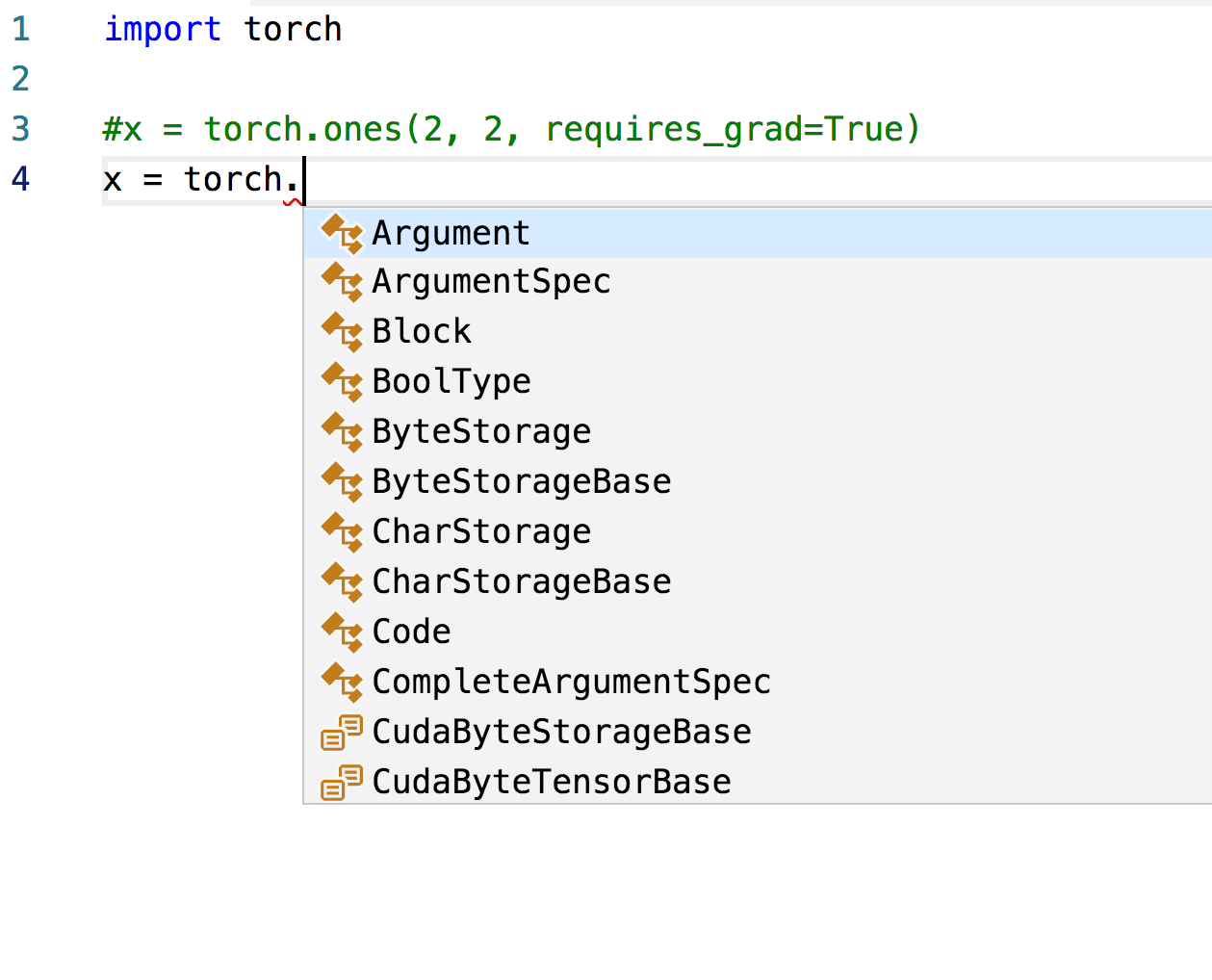}
    \caption{Code completion system providing suggestions in alphabetical order. Left: ``socket'' Python library, right: ``PyTorch''.} 
    \label{fig:example_alphabetic}
\end{figure*}

The most basic code completion system in an IDE would list out all possible
attributes or methods in an alphabetically ordered list. Fig.~\ref{fig:example_alphabetic} shows example completions that such a system could serve. As seen, this approach is capable of yielding accurate results for classes with a relatively small number of members (e.g. ``socket'' Python library), otherwise requiring a developer to scroll through a long list (e.g. in the case of ``PyTorch'').

\subsection{Frequency models}

Instead of ranking methods alphabetically, popularity ranking can be adopted
to increase recommendation relevance. The simplest frequency-based model one could think of would order methods belonging to each class based on the occurrence count in the training corpus. 

Looking for a stronger baseline, we built an improved frequency model incorporating context information about \texttt{if}-conditions (referred to as ``frequency-\texttt{if}'' throughout the paper). In what follows, we subdivide all the invocations into two groups based on whether they are inside an \texttt{if}-condition and creating two corresponding popularity lists. As seen in Tab.~\ref{tab:freq_tf_1}, the most frequent method invocations outside \texttt{if}-statements are related to variable or placeholder declarations, setting up a neural network or training routines; inside \texttt{if}-conditions, however, a typical call would be to file input-output manipulation wrappers, context managers (e.g. eager execution mode), or specifying tensor properties (sparse or dense). As a result, the ``frequency-\texttt{if}'' model would provide more accurate code suggestions inside \texttt{if}-statements increasing overall accuracy.

\begin{table*}[t]
\begin{center}
\begin{tabular}{|c|c|}
\hline
Invocation & Frequency \\
\hline
$\texttt{tf.float32}$              &10508 \\
\hline
$\texttt{tf.nn}$                   &8954 \\
\hline
$\texttt{tf.constant}$             &7754 \\
\hline
$\texttt{tf.train}$                &6869 \\
\hline
$\texttt{tf.variable\_scope}$       &6330 \\
\hline
$\texttt{tf.placeholder}$          &5622 \\
\hline
\end{tabular}
\quad
\begin{tabular}{|c|c|}
\hline
Invocation & Frequency \\
\hline
$\texttt{tf.context.in\_graph\_mode}$                   &905  \\
\hline
$\texttt{tf.gfile.Exists}$                                    &640  \\
\hline
$\texttt{tf.gfile}$                                           &624  \\
\hline
$\texttt{tf.ops.Tensor}$                                      &448 \\
\hline
$\texttt{tf.context.executing\_eagerly}$                       &322 \\
\hline
$\texttt{tf.sparse\_tensor.SparseTensor}$                      &271 \\
\hline
\end{tabular}
\caption{Frequencies of the most popular method calls from the TensorFlow library. Left: outside \texttt{if}-statements, right: inside \texttt{if}-statements.}
\label{tab:freq_tf_1}
\end{center}
\end{table*}

\subsection{Invocation-based Markov Chain model}

Markov Chain models have demonstrated great strength at modeling stochastic transitions, from uncovering sequential patterns \cite{Zimdars2001,Mobasher2002} to
modeling decision processes \cite{Shani2002}. When analyzing invocation occurrences in the source code, we discovered that some sequences of invocations are always followed by the same methods. 

For example, looking at all the function invocations of ``os'' and ``shutil'' Python modules, related to file manipulation, which come after an invocation chain $\texttt{os.path.isfile} \to \texttt{os.remove}$, we have observed that more than 20\% of the time it will be ``os.rename'' or ``shutil.move'', even though those two modules have over 30 methods related to file manipulation.

Considering following three-state Markov process: $$\texttt{os.path.isfile} \to \texttt{os.remove} \to \texttt{?},$$ where the third state can be any function from ``os'' or ``shutil'' modules, we compute the corresponding transition probabilities in Tab.~\ref{fig:three}.
\begin{table*}[t]
\begin{center}
\begin{tabular}{|c|c|}
\hline
Method & Transition probability \\
\hline
\texttt{os.rename, shutil.move} & 0.22 \\ 
\hline
\texttt{os.write} & 0.15 \\
\hline
\texttt{shutil.copy} & 0.13 \\
\hline
\texttt{os.path.isfile} & 0.11 \\
\hline
\texttt{os.remove} & 0.10 \\
\hline
All others methods & 0.29 \\
\hline
\end{tabular}
\end{center}
\caption{Transition probabilities to the third state of the three-state Markov chain: $\texttt{os.path.isfile} \to \texttt{os.remove} \to \texttt{?}$.}
\label{fig:three}
\end{table*}
Using previous invocations occurrences to predict what invocation will be used next is at the core of our Markov Chain model of code completion:
\begin{equation}
P(m_i|h_i) = P(m_i|m_{i-n},..., m_{i-1}), 
\end{equation} 
here, sequence $h_i = m_1, m_2, ..., m_{i-1}$ denotes a history of previous method invocations in the same document scope for a given class. In an $n$-th order Markov chain model the probability of next invocation will only depend on the previous $n-1$ invocations, which is estimated by counting the number of invocation sequence occurrences in the training corpus.  

The abundance of high-quality source code data available on GitHub and the tools allowing to extract and label code snippets are making it possible to train and deploy advanced machine learning models achieving more accurate predictions (cf. section~\ref{sec:neural}).

\section{Dataset}

\begin{figure}
\begin{center}
    \includegraphics[width=.80\textwidth]{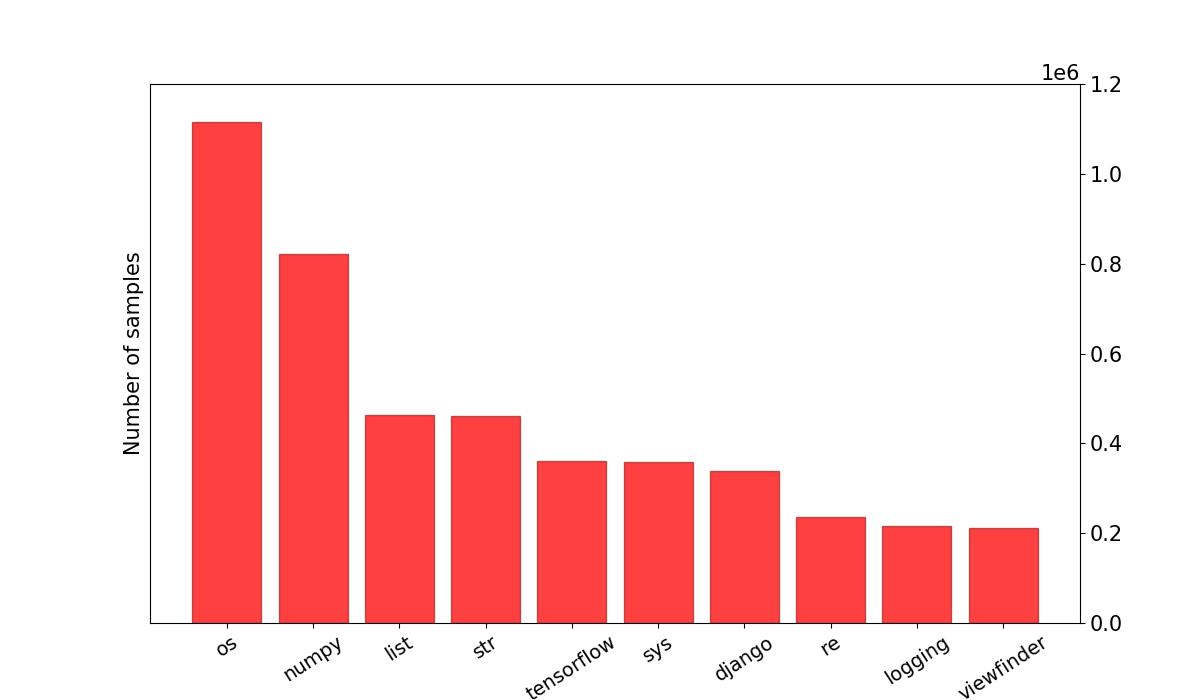}
    \caption{Number of method calls per library, for top 10 libraries in the training dataset.}
    \label{fig:sample_counts}
    \end{center}
\end{figure}

We collected a dataset to train and evaluate code completion models from open source Python repositories on GitHub. A total of 2700 top-starred (non-fork) Python projects have been selected, containing libraries from a diverse set of domains including scientific computing, machine learning, dataflow programming, and web development, with over 15.8 million method calls. Fig.~\ref{fig:sample_counts} shows Python libraries with the most method call occurrences in the dataset.

We split the dataset into development and test sets in the proportion 70-30 on the repository level. The development set is then split at random into training and validation sets in the proportion 80-20. To serve predictions online, the model is retrained using the entire dataset.

\section{Representing code snippets}
\label{sec:code2vec}

Any programming language has an explicit context-free grammar (CFG), which can be used to parse source code into an AST. An AST is a rooted n-ary tree, where each non-leaf node corresponds to a non-terminal in the CFG specifying structural information. Each leaf node corresponds to a syntax token encoding program text. 

The Pythia model consumes partial file-level ASTs corresponding to code snippets containing member access expressions and module function invocations as an input for training. Before feeding to LSTM, ASTs are serialized to sequences according to in-order depth-first traversal. We retain up to $T$ lookback tokens preceding each method call when extracting training sequences, where $T$ is a tunable parameter of the model provided in~Tab.~\ref{tab:example_hyperparameters}. 

Further, in order for an AST to be consumable by LSTM, its nodes and tokens have to be mapped to numeric vectors. Like the weights of the LSTM, the look-up tables mapping discrete values of tokens to real vectors can be learned via backpropagation. The Word2Vec approach~\cite{NIPS2013_5021} has been shown to work well in Natural Language Processing, and is employed in the Pythia system to represent code snippets as dense low-dimensional vectors.

\begin{figure*}
    \includegraphics[width=.95\textwidth]{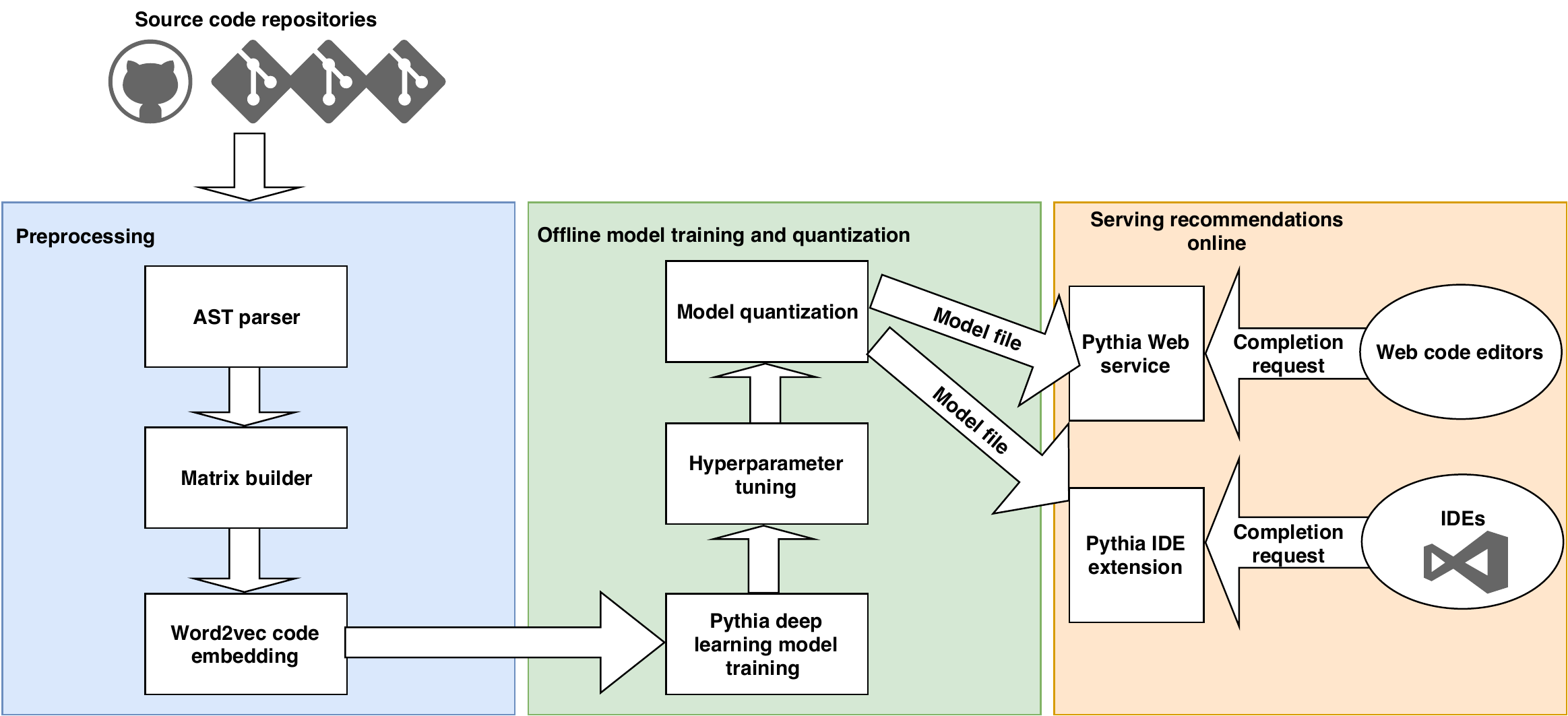}
    \caption{Pythia workflow: from raw source code files to serving recommendations online.}
    \label{example_workflow}
\end{figure*}

Each input data file is parsed using the PTVS parser~\footnote{https://github.com/Microsoft/PTVS} to extract its AST. The method invocations and associated metadata, including invocation spans, serialized sequences of syntax nodes and tokens from the AST, and the runtime types of method call receiver tokens are extracted. The data processing workflow is summarized in Fig.~\ref{example_workflow}.

Syntax node and token names are mapped to integers from 1 to $V$ to obtain token indices. Infrequent tokens are removed to reduce vocabulary size. During training, the out-of-vocabulary (OoV) tokens in a code snippet are mapped to integers greater than vocabulary size $V$, in a way that the occurrences of the same OoV token in the training code snippet would be represented by the same integer. Each training sample is an AST serialized into sequence terminated with the "." end-of-sequence character. A name of the method call token serves as a label and is one-hot encoded. We train a word embedding on the corpus of the source code repositories to map token indices to low-dimensional dense vectors preserving semantic relationships in the code.

\subsection{Leveraging type information}

Python is a dynamically typed language. While PEP 484\footnote{https://www.python.org/dev/peps/pep-0484} introduces type annotations for various common types starting Python 3.6, most of the libraries have not yet adopted it. As such, Python interpreter does variable type checking only as code runs. We infer types of the method call receiver tokens and local variables at runtime based on static analysis of the object usage patterns and include this information into training sequences.

In Python programming language, a user has freedom of aliasing imported modules. For instance, a commonly used ``import numpy as np'' may well be ``import numpy as an\_unusual\_alias''. Embedding vectors for ``np'' and ``numpy'' tokens are nearby in embedding space. However, an infrequent alias like ``an\_unusual\_alias'' may be an out-of-vocabulary token and not represented in the embedding. Including type information during training means an embedding vector corresponding to ``numpy'' token would be used in all of the cases described above.

Variable names are nouns chosen by a developer, which may differ from one code implementation to another. Keeping them in the original format increases the vocabulary size, making a code completion system dependent on spelling of a variable name. To avoid this, variable names are normalized according to $var:<variable type>$ convention.  

\section{Neural code completion model}
\label{sec:neural}

Recurrent Neural Networks are a family of neural networks capable of processing sequential data of arbitrary length. Long short-term memory networks \cite{lstm}, a particularly successful form of RNNs, trained on large datasets can obtain remarkable performance results across a wide variety of domains -- from image captioning~\cite{mao2014deep}, to sentiment analysis and machine translation~\cite{Socher-etal:2013,NIPS2014_5346}. An LSTM unit uses no activation function within its recurrent components, controlling information flow using gates implemented using the logistic function. Thus, the stored values are not iteratively expanded or squeezed over time, and the gradient does not tend to explode or vanish when trained. 

The task of method completion is to predict a token $m^{*}$, conditional
on a sequence of syntax tokens ${c_{t}}, t=0...T$, corresponding to the terminal nodes of the AST of a code snippet $C$, plus the special end-of-sequence token ".". This problem is a natural fit for sequence learning with LSTM:
\begin{eqnarray}
x_t &=& Lc_t,\\ 
h_t &=& f(x_t, h_{t-1}),\\ 
P(m|C) &=& y_t = softmax(Wh_t+b), \\
m^{*} &=& argmax(P(m|C)).
\end{eqnarray}
Here, the matrix $L \in R^{d_x\times|V|}$ is the word embedding matrix, $d_x$ is the word embedding dimension and $|V|$ is the size of the vocabulary. In case of Pythia, function $f(.,.)$ represents a stacked LSTM taking the current input and the previous hidden state and producing the hidden state at the next temporal step. $W \in R^{|V|\times d_h}$ and $b\in R^{|V|}$ are the output projection matrix and the bias. The $d_h$ is the size of the hidden state of LSTM.  

\begin{figure*}
    \includegraphics[width=.95\textwidth]{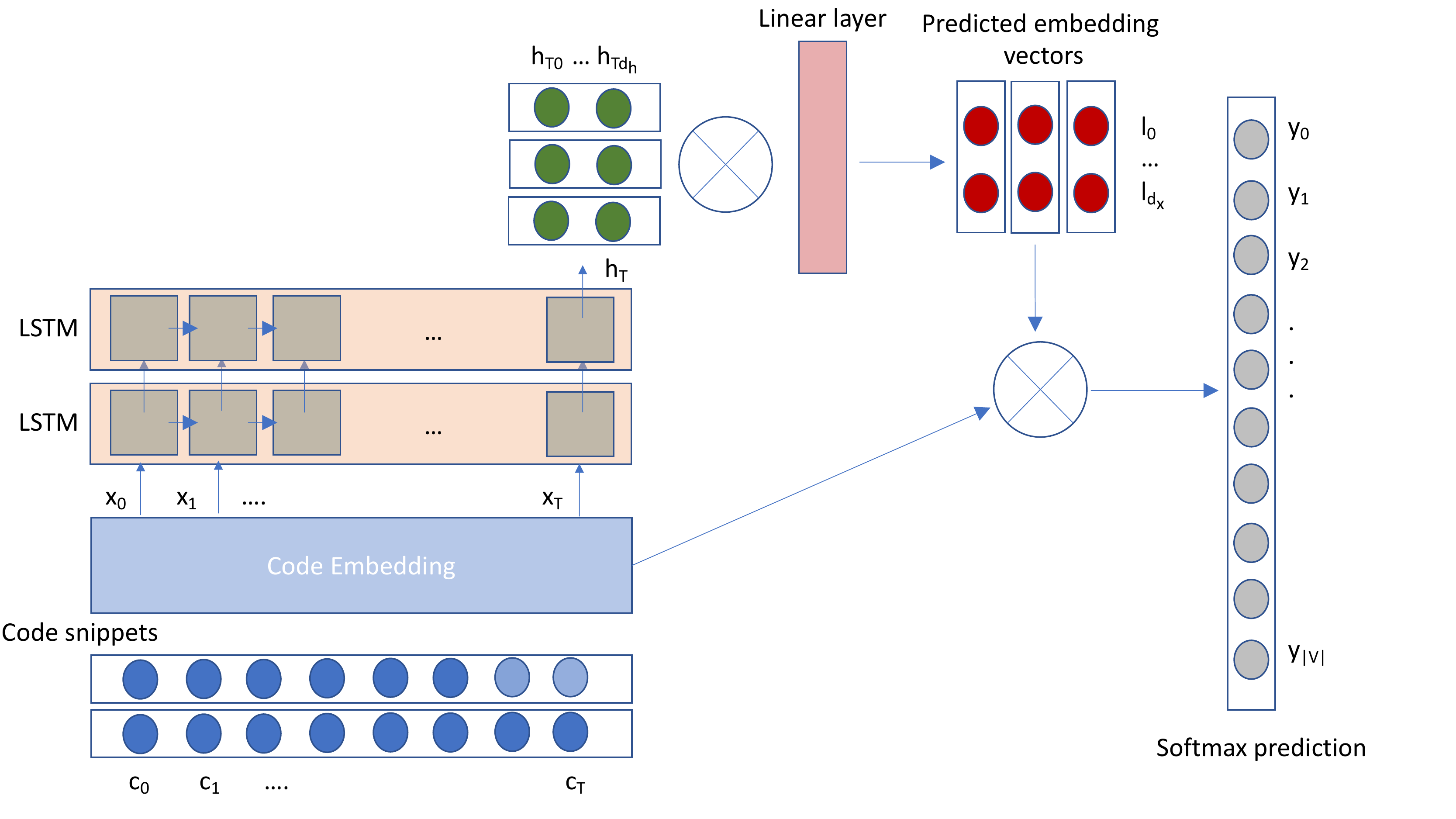}
\caption{Architecture of the neural network deployed in the Pythia code completion system.}
\label{fig:NNcartoon}
\end{figure*}

Inspired by~\cite{DBLP:journals/corr/InanKS16}, we are reusing the input word embedding matrix as the output classification matrix as shown in Fig.~\ref{fig:NNcartoon}, which allows to remove the large fully connected layer and significantly reduce the number of trainable parameters and the model size on disk.
More specifically, we introduce a projection matrix $A = (a)_{ij} \in R^{d_{h}\times d_{x}}$ initialized according to a random uniform distribution. Given an LSTM encoded snippet and a hidden state at the last temporal step $h_T \in R^{d_{h}}$, by multiplying the two together we obtain the predicted embedding vector $L^{pred} = (l^{pred})_{j} \in R^{d_{x}}$ as:
\begin{equation}
l^{pred}_{j} = \sum_{i} h_{Ti}a_{ij}.
\end{equation}
Subsequently, the unnormalized predictions of the neural network are obtained as:
\begin{equation}
y_k = \sum_{j} l_{kj}l^{pred}_{j} + b_{k} 
\end{equation}
where ${b}_{k}, k=0...|V|-1$ is the bias vector initialized to zeros before backpropagation.

\section{Model training}

Neural networks are trained iteratively, making multiple passes over an entire dataset before converging to a minimum. Backpropagation through time (BPTT) is a gradient-based neural network training algorithm we apply to train the LSTMs statefully. 

Training deep neural networks is a computationally intensive problem that requires the engagement of high-performance computing clusters. Pythia uses a data-parallel distributed training algorithm with Adam optimizer, keeping a copy of an entire neural model on each worker, processing different mini-batches of the training dataset in parallel lockstep.

The offline training module of the Pythia system is implemented as a Python library integrating TensorFlow and CUDA-aware MPI for distributed training. The software stack makes use of CUDA 9, GPU accelerated deep learning primitives from CuDNN 7, and TensorFlow 1.10. We use BatchAI~\footnote{https://docs.microsoft.com/en-us/azure/batch-ai/overview} -- an Azure cloud service providing on-demand GPU clusters with Kubernetes resource manager -- for model training and hyperparameter optimization. The online module is implemented in C\#, making use of ML.NET library and ONNX data format~\footnote{https://github.com/onnx/onnx}. 

\subsection{Batching and Learning rate schedule}

The number of nodes in a Python file-level abstract syntax tree ranges from $O(10^2)$ to $O(10^4)$. To leverage long-range dependencies in the source code we consider sequence lengths in the range of $100 - 1000$. To overcome the gradient vanishing problem for long sequences, we approximate the computation of the gradient of the loss with respect to the model parameters by truncated backpropagation through time~\cite{DBLP:journals/corr/Graves13}.

\begin{figure*}
    \includegraphics[width=.95\textwidth]{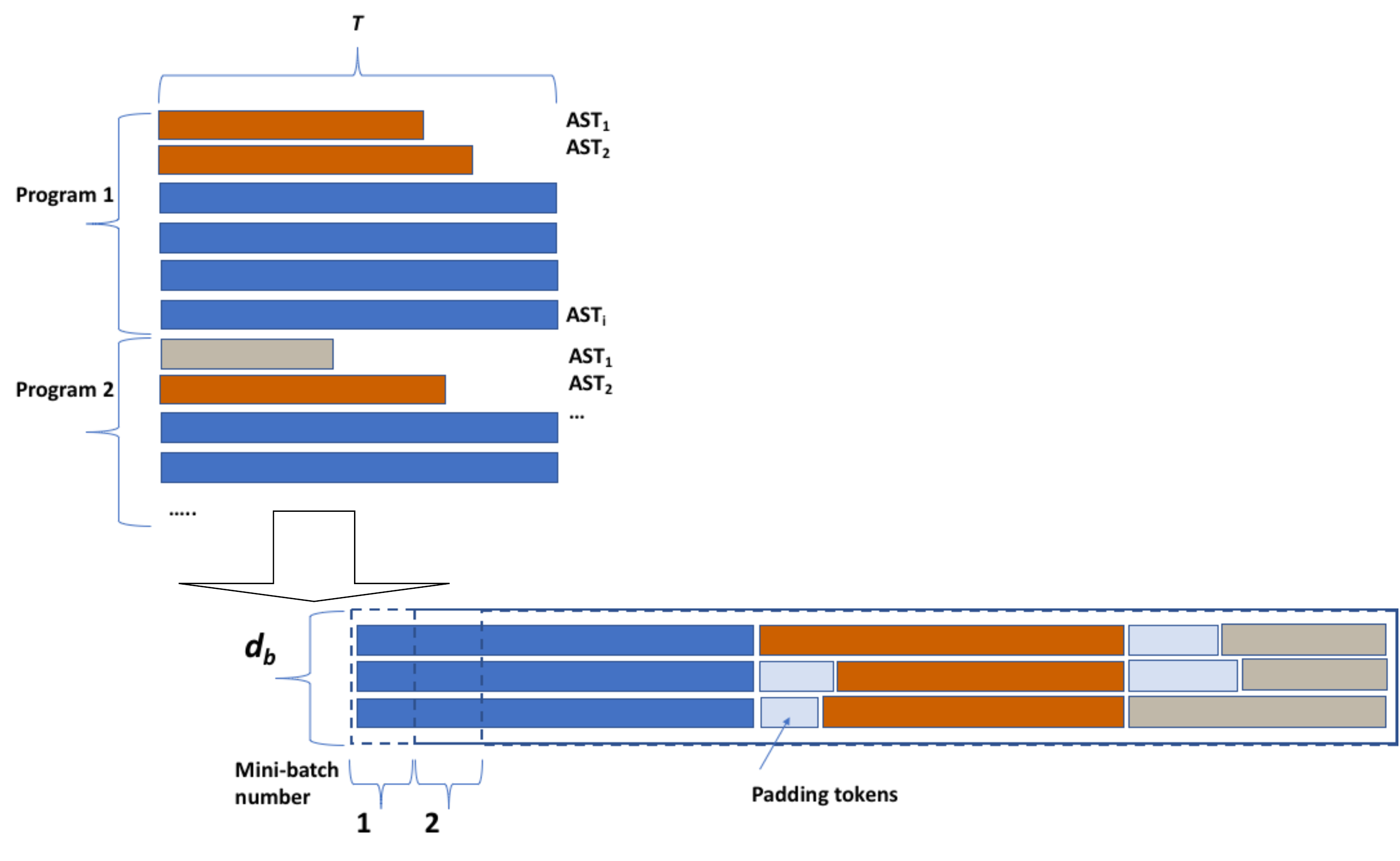}
\caption{Snapshot of the training buffer.}
\label{fig:train_seqs}
\end{figure*}

Efficient batching is important to fully utilize parallelism of modern GPUs. In general, training sequences will have variable lengths, with a majority having the maximum length of $T$, as shown in Fig.~\ref{fig:train_seqs}. Training samples are presorted and split into three buckets based on their lengths. Within each bucket, sequences are padded to the maximum length using the special padding token, which is then excluded from loss calculation by means of masking. A training buffer to maintain sequences belonging to $d_{b}$ distinct ASTs is allocated. At every training step, the first $T_{RNN}$ timesteps of the buffer are fed as the next batch. The buffer is then shifted by $T_{RNN}$. Every time a sequence is finished in the buffer, a new set of sequences is loaded, and the LSTM internal states are reset for training.  


The learning rate controlling the magnitude of the weight update during gradient optimization is lowered upon completion of each epoch according to the exponential decay. In a distributed regime, the learning rate is scaled up during the first few epochs to facilitate reliable model convergence. Closely following~\cite{facebook_paper}, we linearly scale the learning rate up proportionally to the number of workers $N_{worker}$ during the first 4 epochs (``warm-up'' period):
\begin{equation}
\lambda_{0}(N_{worker}) = \lambda_{0}\cdot \gamma^{i}\cdot \frac{N_{worker}}{\alpha}
\end{equation}
here, $\lambda_{0}$ is the base learning rate, $\gamma$ is the learning rate decay constant, $i$ is the epoch number, and parameter $\alpha$ is the scaling fraction controlling the learning rate at the end of the warm-up period, $\alpha=4$ was found to work best via hyperparameter search.

\begin{figure*}
    \includegraphics[width=.49\textwidth]{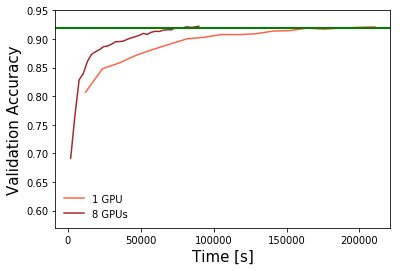}
    \includegraphics[width=.49\textwidth]{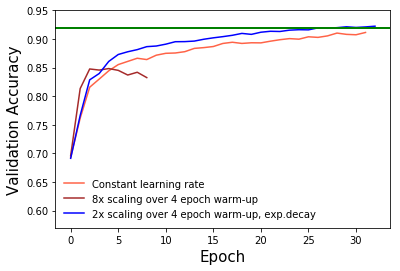}
    \caption{Accuracy on the validation set for various training regimes. Left: serial and distributed training with 8 worker GPUs; right: distributed training with 8 GPUs for various learning rate schedules. The horizontal green line indicates the target model accuracy of 92\%.}
    \label{accs}
\end{figure*}

Fig.~\ref{accs} shows the reduction in training time due to distributed training and the effect of learning rate schedule.  

\subsection{Hyperparameters}

Overall, the model architecture, training procedure, and data normalization
produce a large number of hyperparameters that must be tuned to maximize predictive performance. These hyperparameters include numerical values such as the learning rate and number of LSTM layers, dimension of embedding space, but also abstract categorical variables such as the precise model architecture or the normalization algorithm. These parameters are summarized in Tab.~\ref{tab:example_hyperparameters}.

\begin{table*}[t]
\begin{center}
\begin{tabular}{|c|c|c|}
\hline
Hyperparameter & Explanation & Best value \\
\hline
$\lambda_{0}$ & Base learning rate & 0.002 \\
\hline
$\gamma$ & Learning rate decay per epoch & 0.97 \\
\hline
$N$ & Number of recurrent neural network layers & 2 \\
\hline
$d_h$ & Number of hidden units in LSTM, per layer & 100 \\
\hline
$T$ & Number of lookback tokens, & 1000 \\
& timesteps through which backpropagation is run &  \\
\hline
$T_{RNN}$ & Number of timesteps through which & 100 \\
& backpropagation is run & \\
\hline
RNN type & Type of RNN & LSTM \\
\hline
$d_{b}$ & Batch size & 256 \\
\hline
Loss function & Type of loss function & Categorical cross-entropy\\
\hline
$d_{x}$ & Embedded vector dimension & 150 \\
\hline
Optimizer & Stochastic optimization scheme & Adam \\
\hline
Dropout & Dropout keep probability & 0.8 \\
\hline
L2 Regularization & Weight regularization of all layers & 10 \\
\hline
Clip norm & Maximum norm of gradients & 10 \\
\hline
Token frequency threshold & Minimum frequency of syntax token in the corpus & 500 \\
& for inclusion in vocabulary & \\
\hline
\end{tabular}
\end{center}
\caption{Hyperparameters to be optimized together with explanations and
well-performing values.}
\label{tab:example_hyperparameters}
\end{table*}
Throughout this work, the ``best'' model is determined by hyperparameter
tuning. This is done via random search in the respective hyperparameter space of each method, i.e. by training a number of models with random hyperparameters on the training set and choosing the one with the highest performance on validation set.

\subsection{Tuning neural network architecture}

Architecture of the neural network is a hyperparameter. Besides LSTM, we consider gated recurrent units (GRU), LSTM with attention mechanism for temporal data~\cite{DBLP:journals/corr/RaffelE15}, and variations of classification layer following the LSTM.

Selecting the best model for serving online is a trade-off between accuracy and model size. Tab.~\ref{tab:nn_tune} shows validation level accuracy for various neural model architectures and associated sizes on disk.
\begin{table*}[t]
\begin{center}
\begin{tabular}{|c|c|c|c|}
\hline
Model architecture & Top-5 accuracy & Model size (quantized size), MB \\
\hline
LSTM+fully connected & 0.91 & 202 (51) \\ 
\hline
GRU+predicted embedding & 0.91  & 152 (38) \\ 
\hline
LSTM+predicted embedding & 0.92 & 152 (38) \\
\hline
LSTM+attention & 0.93 & 164 (41) \\ 
\hline
\end{tabular}
\end{center}
\caption{Accuracy on the validation set for various neural model architectures and associated model sizes.}
\label{tab:nn_tune}
\end{table*}
As seen, removing a large fully connected classification layer in favor of predicted embedding reduces the number of trainable parameters and the model size on disk by 25\%. The best top-5 accuracy is achieved when the attention mechanism is employed, however, this model is 8\% larger and has slower inference speeds. Consequently, we have chosen the LSTM with predicted embedding for deployment.

\section{Evaluation}
\label{sec:evaluation}


\subsection{Evaluation Metrics}

Top-k accuracy and the mean reciprocal rank (MRR)~\cite{Radev2002} are
used to measure the quality of recommendations, defined as:
\begin{eqnarray}
Acc(k) &=& \frac{N_{top-k}}{Q}, \\   
MRR &=& \frac{1}{Q}\cdot\sum_{i=1}^{Q}\frac{1}{rank_i},    
\end{eqnarray}
where $N_{top-k}$ denotes the number of relevant recommendation in top $k$ suggestions, $Q$ represents the total number of test data samples and $rank_i$ is the prediction rank of a recommendation.

Accuracy in top-1 tells how often the top recommendation is correct, while top-5 accuracy provides
an idea of how often the top five recommendation list contains the suggestion a user is looking for. The MRR captures the overall rank of the result, thus providing information of how far outside of the list of top suggestions the model prediction was. MRR values closer to one indicate overall smaller recommendation ranks, corresponding to a better performing model.

\subsection{Evaluation results}

\begin{figure}
\begin{center}
    \includegraphics[width=.70\textwidth]{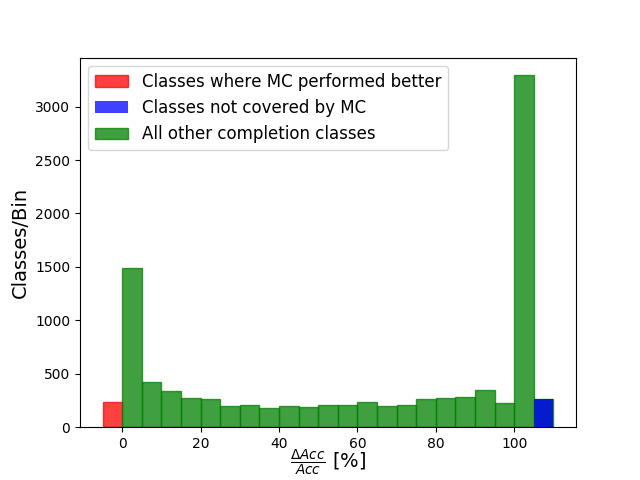}
    \caption{Number of completion classes with a given relative percentage accuracy difference for the Pythia neural model as compared to the invocation-based Markov Chain baseline. $\frac{\Delta Acc}{Acc} = 100\cdot(Acc_{Pythia}-Acc_{MC})/Acc_{MC}$ [\%]}
    \label{example_improve}
    \end{center}
\end{figure}

Tab.~\ref{tab:basic_comparison} shows the performance comparison of the Pythia neural model and various simpler baselines, including basic alphabetic ordering, frequency models, and invocation-based Markov Chain model.
\begin{table*}[t]
\begin{center}
\begin{tabular}{|c|c|c|c|}
\hline
Method & Top-1 accuracy& Top-5 accuracy & MRR \\
\hline
Alphabetic & 0.36 & 0.47 & 0.372 \\ 
\hline
Frequency & 0.38 & 0.64 & 0.495 \\ 
\hline
Frequency-if & 0.40 & 0.67 &  0.521 \\ 
\hline
Markov Chain & 0.58 & 0.83 & 0.704 \\
\hline
Pythia & 0.71 & 0.92 & 0.814\\
\hline
\end{tabular}
\end{center}
\caption{Accuracy and mean reciprocal ranks on the test set for the Pythia neural model and various baselines.}
\label{tab:basic_comparison}
\end{table*}
As seen, our model significantly outperforms all the baselines, especially for the accuracy of top-1 suggestions. Tab.~\ref{tab:improvements} summarizes the performance of the Pythia system for 10 most popular completion classes.
\begin{table*}[t]
\begin{center}
\begin{tabular}{|c|c|c|c|c|}
\hline
Class name & Top-5 accuracy, MC & Top-5 accuracy, Pythia & $\frac{\Delta Acc}{Acc}~[\%]$ & $N_{recommend}$\\
\hline
os                     &0.863              &0.950   &10.0 &125673 \\    
\hline
numpy                  &0.575     &0.697   &21.2 &113227  \\    
\hline
list                   &0.978            &0.989   &1.1 &79406  \\    
\hline
str                    &0.974            &0.988    &1.4 &77507 \\     
\hline
os.path                &0.895            &0.957    &6.8  &75270  \\    
\hline
sys                    &0.821            &0.959   &16.8  &51231  \\    
\hline
wx                     &0.272            &0.533   &95.9  &47821 \\     
\hline
logging                &0.846            &0.914   &8.01  &40871 \\    
\hline
time                   &0.951            &0.980   &3.1  &27874  \\     
\hline
tensorflow             &0.511             &0.754   &47.6  &16620 \\ 
\hline
\end{tabular}
\end{center}
\caption{Performance summary on the test set for the Pythia neural model and the Markov Chain baseline. Relative percentage accuracy change is defined as: $\frac{\Delta Acc}{Acc} = 100\cdot(Acc_{Pythia}-Acc_{MC})/Acc_{MC}$ [\%]}
\label{tab:improvements}
\end{table*}
Fig.~\ref{example_improve} illustrates a performance improvement achieved by the Pythia neural model as compared to the Markov Chain baseline on the test set. As seen from the histogram, Pythia yields over 50\% accuracy improvement for nearly 6000 completion classes. Markov Chain model relies on the type inference, which results in a lower coverage than the Pythia neural model. The classes not covered by the Markov Chain model are included in the overflow bin.

\section{Model deployment}
\label{sec:deploy}

The Pythia code completion system is currently deployed as a part of Intellicode extension~\footnote{https://marketplace.visualstudio.com/items?itemName=VisualStudioExptTeam.vscodeintellicode} in Visual Studio Code IDE. The main challenges in designing and implementing an online system capable of serving models on lightweight client devices are the prediction latency and memory footprint. We apply neural network quantization into an 8-bit unsigned integer numeric format to reduce memory footprint and model size on disk. 

\subsection{Neural network quantization}

Neural network quantization is a process of reducing the number of bits to store weights. The numerical format used during training of the Pythia neural model is IEEE 754 32-bit float. Quantization is performed layer-by-layer, extracting minimum and maximum values of weights and activations in the floating point format, zero shifting, and scaling. Given a weight matrix $W = {w}_{ij}$ for a layer, the quantized weight matrix $W^{q}$ is obtained as:
\begin{equation}
\beta = \frac{max(W) - min(W)}{2^{8}}, \quad
w^{q}_{ij} = \frac{w_{ij} - min(W)}{\beta}.  
\end{equation}

In the case of Pythia, post-training quantization into 8-bit integer results in model size reduction to quarter the size -- from 152 MB to 38 MB -- reducing top-5 accuracy from 92\% to an acceptable 89\%. 

\begin{figure}
\begin{center}
    \includegraphics[width=.70\textwidth]{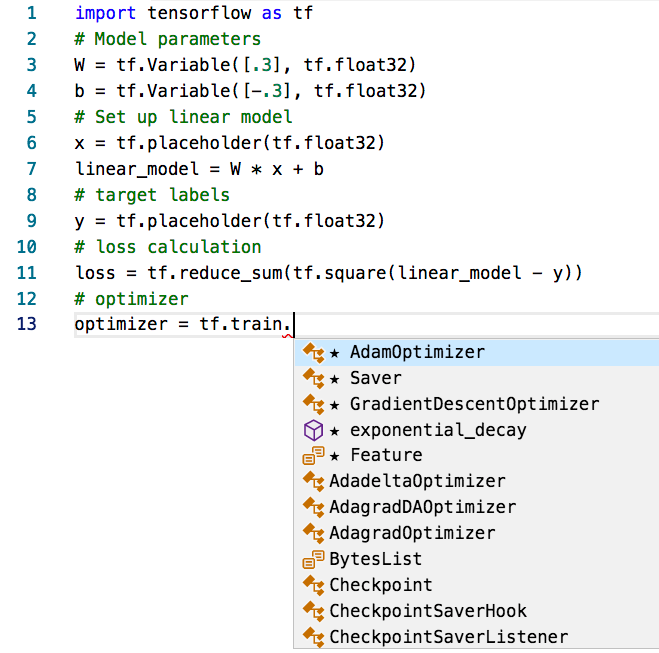}
    \caption{Example code completion served by Pythia.} 
    \label{example_ast}
\end{center}
\end{figure}

An example online recommendation served by the Pythia neural model is shown in Fig.~\ref{example_ast}.
As seen, Pythia accurately suggests the most likely method call should involve an optimizer, with two out of the top five recommendations being Adam and SGD optimizers. Other suggestions include ``Saver'' -- a method adding ops to save and restore variables to and from checkpoints, ``exponential\_decay'' -- controlling learning rate decay, and ``Feature'' -- a data wrapper specific to TensorFlow, all of which are likely to be called to before creating a session and running the training ops.    

\section{Conclusions}

We have introduced a novel end-to-end approach for AI-assisted code completion called Pythia, generating ranked lists of method and API recommendations that can be used by software developers. We have deployed the system as part of Intellicode extension in Visual Studio Code IDE. Pythia makes use of long short-term memory networks trained on long-range code contexts extracted from abstract syntax trees, capturing semantics carried by distant nodes. Evaluation results on a large dataset of 15.8 million method calls extracted from real-world source code showed that our best model achieves 92\% top-5 accuracy, beating simpler baselines. 

We have overcome and documented several practical challenges of training deep neural networks and hyperparameter tuning on high-performance computing clusters, and model deployment on lightweight client devices to predict the best matching code completions at edit time.

Besides Python, Intellicode extension is providing AI-assisted code completions for a variety of programming languages including C\#, Java, C++, and XAML based on our Markov Chain model. In the future, advanced deep learning approaches will be explored in application to programming languages other than Python, and for more sophisticated code completion scenarios.

%
\section{Acknowledgement}
We are thankful to Miltiadis Allamanis of Microsoft Research for valuable discussions on the neural network architectures and reduction of number of trainable parameters. We also thank Microsoft AI Frameworks and ML.NET teams for helping deploying the model, as well as to Christian Bird of Microsoft Research for reading the manuscript.

\bibliographystyle{unsrt}  
\bibliography{references}  

\end{document}